\documentclass[cameraready]{Interspeech}

\title{A Toolkit for Detecting Spurious Correlations in Speech Datasets}

\author[affiliation={1,2}]{Lara}{Gauder}
\author[affiliation={1}]{Pablo}{Riera}
\author[affiliation={3}]{Andrea}{Slachevsky}
\author[affiliation={3}]{Gonzalo}{Forno}
\author[affiliation={4}]{Adolfo M.}{García}
\author[affiliation={1}]{Luciana}{Ferrer}

\address{
    $^1$ Instituto de Investigación en Ciencias de la Computación, UBA-CONICET, Argentina \\
    $^2$ Departamento de Computación, Facultad de Ciencias Exactas y Naturales, UBA, Argentina \\
    $^3$ Facultad de Medicina, Universidad de Chile, Chile \\
    $^4$ Centro de Neurociencias Cognitivas, Universidad de San Andrés, Argentina
}

\email{mgauder@dc.uba.ar, priera@dc.uba.ar, lferrer@dc.uba.ar, adolfo.garcia@gbhi.org}


\usepackage{comment}

\begin{document}

\maketitle

\begin{abstract}
We introduce a toolkit for uncovering spurious correlations between recording characteristics and target class in speech datasets. Spurious correlations may arise due to heterogeneous recording conditions, a common scenario for health-related datasets. When present both in the training and test data, these correlations result in an overestimation of the system performance -- a dangerous situation, specially in high-stakes application where systems are required to satisfy minimum performance requirements. Our toolkit implements a diagnostic method based on the detection of the target class using only the non-speech regions in the audio. Better than chance performance at this task indicates that information about the target class can be extracted from the non-speech regions, flagging the presence of spurious correlations. The toolkit is publicly available for research use.
\end{abstract}

\section{Introduction}

Spurious correlations are statistical associations between input features and the target variable that arise from dataset-specific biases rather than from a genuine relationship relevant to the prediction task~\cite{geirhos2020shortcut, steinmann2024navigating, sahidullah2025shortcut, pathak.2025}. Models trained on such data may learn to predict the target class using irrelevant features from the data~\cite{jamadermatol.2019, sagawa2019distributionally, degrave2021ai, huang2022developing}. 
Crucially, this problem cannot be diagnosed by evaluating performance on samples extracted from the same collection, since the learned short-cut would still be useful on that data. A decision may then be made to adopt the system for a given application based on its superior performance on this dataset. 
Yet, given a new scenario where the spurious correlation is not present, the system would unpredictably fail. 

In speech datasets, spurious correlations may arise due to problems in the collection protocol. 
For example, health-related data collections often occur in more than one location (some examples of datasets that were collected in more than one location include \cite{jesus2017advanced, lopes2024automated, sanz2022automated, braun2024infusing}). Even if the same recording device is used in all locations, the background noise, room reverberation characteristics, or distance to the microphone will likely differ. Hence, if care is not taken to ensure that the patients with the condition of interest and control subjects are equally represented in each location, a correlation between the target class (patient vs control) and the audio characteristics will exist. When using such data for training, models relying on low-level acoustic features or embeddings from self-supervised models which are able to encode such information,  may learn to detect the type of background noise as a short-cut to help detect the class. 


For some datasets, the metadata (like the recording location or device) may provide the information required to uncover a spurious correlation.  
In some cases, though, the confounding factor that correlates to the target class is not part of the metadata. For example, doctors recording a patient with the condition of interest may type the patient's responses during the session, while doctors recording a control subject may not. The presence of the typing noises would introduce a spurious correlation that may easily go unnoticed.
Crucially, as we will see in this work, while audio enhancement techniques can be used to improve the perceptual quality of audio samples, models trained on enhanced audio may still be able to detect a difference in conditions from the traces left by the enhancement process.

Motivated by these challenges, we developed a toolkit to diagnose whether a speech dataset contains spurious correlations related to background noise, microphone characteristics, encoding method, reverberation, or other recording aspects that affect the full waveform both over speech and non-speech regions. The method is specifically designed for datasets for which the following conditions are satisfied: 1) the waveforms contain at least a few seconds of silence, 2) the labels of interest are assigned at the waveform level and describe speech-related characteristics, and 3) a binary classification task can be derived from the labels (for multiclass tasks, one-versus-other subtasks can be defined for the analysis; for regression, the labels can be discretized). This covers many speech processing problems, including medical tasks like Alzheimer disease or voice pathology detection, as well as other waveform-level tasks like emotion, speaker or language classification. 

The method relies on the assumption that, for a task that is defined by the speech characteristics, the non-speech parts of the signal should carry no information about the target class, except, perhaps, through timing information. 
The proposal is then to learn a model to predict the target class using only the non-speech regions of the signal. If the model gives a better-than-chance performance for this task, then it must be learning to predict the class based on the acoustic characteristics of the recording rather than on the subject's characteristics. Importantly, though, great care needs to be taken to avoid inadvertently modeling speech characteristics through the analysis of the non-speech regions, since the timing of such regions may be useful to detect the target class. Further, leakage of speech sounds due to errors in voice-activity detection (VAD), may lead to incorrect conclusions. While an approach conceptually similar to ours was proposed in a prior work \cite{liu2024cleverhans}, their method used uncurated VAD outputs and was based on  features and models that  could easily leverage timing information. The paper diagnosed a problem with ADReSS$_o$, a commonly used Alzheimer dataset. Our results suggest that such diagnosis partly stemmed from leaking of speech information into the model. 

In our toolkit, we take the following steps to avoid misdiagnosing spurious correlations. First, we provide tools for selecting the voice-activity-detection system, either by auditing the regions annotated as non-speech, or by validation using a small subset of manually annotated data. Second, we use short-term local features paired with a careful modeling approach that is intrinsically unable to encode timing information. 

Our publicly available toolkit\footnote{\url{https://github.com/habla-liaa/spurious-correlation-detection-toolkit}} allows researchers to easily perform this important sanity check on their own datasets. We argue that this type of analysis should always be done before drawing conclusions from speech-based experiments, specially when using datasets recorded in several locations or under uncontrolled conditions. 

\section{Proposed Method} \label{sec:methods}

\begin{figure}[t]
    \centering
    \includegraphics[scale=0.3]{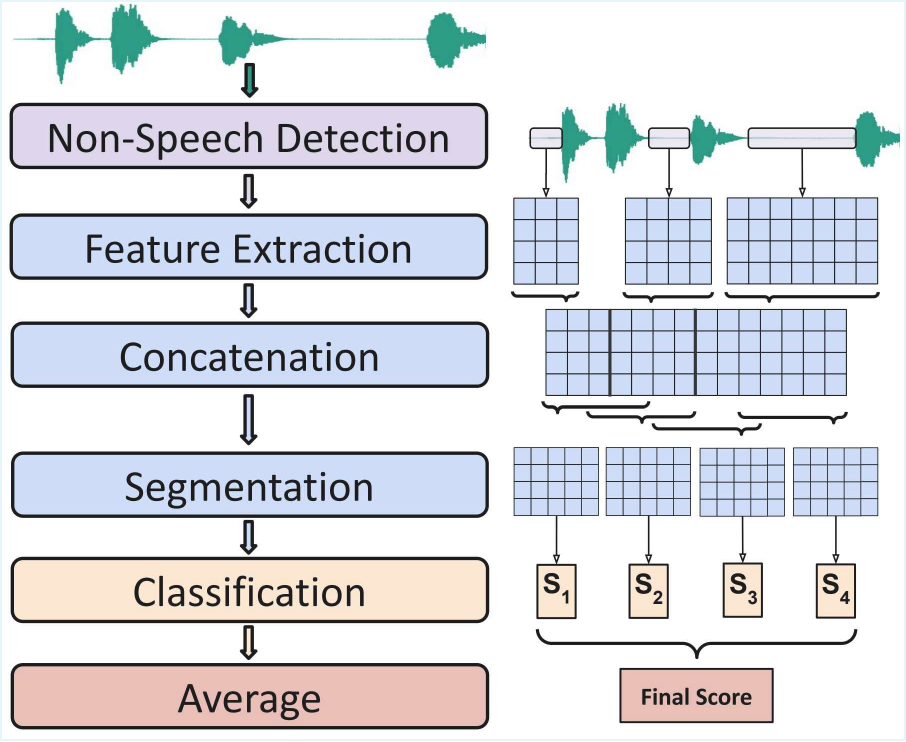}
    \caption{Schematic of the proposed method. Features are extracted over the non-speech segments and concatenated into a single sequence, which is then split into overlapping chunks. Scores are obtained with a classifier trained with chunks obtained from the training data. Finally, the scores over all chunks are averaged to produce the score for the input sample.}
    \vspace{-0.35cm}
    \label{fig:pipeline}
\end{figure}

In this section, we describe the proposed method for uncovering the presence of acoustic spurious correlations in a given dataset. 
The pipeline involves several steps. First, the non-speech parts are extracted from the signals, either with a voice-activity detection system (VAD) or using manual annotations. Second, acoustic features are extracted over each non-speech segment, concatenated, and segmented into fixed-length chunks. The chunks are then used for training a system to detect the class of the sample from which the chunk was extracted. Then, the system is run over all test chunks and the scores for all chunks within a waveform are averaged. Finally, the discrimination performance of those scores is calculated. If the performance is significantly above chance, then it should be concluded that the recording conditions of the samples correlate with the target class.
The pipeline used for inference is illustrated in Figure \ref{fig:pipeline}.


\subsection{Voice Activity Detection (VAD)} \label{sec:vad}
Accurate detection of the non-speech regions is a critical part of our approach. Leakage of speech sounds into the regions used for the analysis could result in the incorrect conclusion that spurious correlations exist when, in fact, the model is simply using the speech information that was mislabeled by the VAD system. Hence, ideally, manual annotations of non-speech regions should be used. When obtaining manual alignments for all recordings is infeasible, an alternative procedure is to annotate a few samples from the dataset and use them as development data for selecting the best VAD system for the type of condition present in the data.

The toolkit supports several VAD systems (Pyannote \cite{Plaquet23}, Silero \cite{Silero}, Whisper \cite{radford2023robust}, TorchVAD \cite{torchaudio}, and SpeechBrain \cite{speechbrain}), which can be run to obtain a segmentation into speech/non-speech regions. Since some VAD systems normalize the audio before processing, relatively low-volume speech may be labeled as non-speech. To mitigate this problem which would result in speech information leaking into the non-speech regions, the toolkit allows for a second round of VAD to be run on the non-speech regions from the first round. 


To assess the quality of the VAD system, the toolkit computes two metrics: speech leakage (the fraction of true speech frames erroneously labeled as non-speech averaged over samples) and missed non-speech (the fraction of non-speech frames incorrectly labeled as speech averaged over samples). A tool to visually compare one or more VAD outputs with the ground truth on a given sample is provided in the toolkit.

Finally, the toolkit also provides an interactive script to listen to the non-speech segments produced by a given VAD configuration and annotate potential problems (e.g., residual speech or excessive noise). 
For cases in which manual annotation is not feasible, we recommend running the Silero VAD with a conservative threshold for detecting speech (e.g., 0.2), running the second VAD stage, and using this auditing tool to listen to all the audio detected as non-speech. If speech is found in a few samples, they can be discarded from the subsequent stages of the process. If speech is found in too many of the samples, a different VAD system or threshold may need to be considered.

\subsection{Speech Enhancement} \label{sec:speech_enhancement}
It is reasonable to assume that spurious correlations due to background noise may be mitigated by denoising approaches. Our toolkit implements a speech enhancement pipeline consisting of a first stage of loudness normalization based on the EBU R128 standard \cite{ebu2011loudness} followed by filtering with the DeepFilterNet model \cite{schroeter2022deepfilternet}, which is designed to suppress stationary ambient noise and certain transient artifacts. Audio examples demonstrating the impact of this speech enhancement tool are available in the toolkit’s published repository.

\subsection{Acoustic Feature Extraction} \label{sec:features}
Prior work that proposes an approach similar to ours \cite{liu2024cleverhans} uses wav2vec 2.0 (W2V2, for short) \cite{baevski2020wav2vec} as feature extractor. A potential problem with using these features for our purposes is that they may be able to encode information about the duration of the segments from which they are extracted, since the model is a transformer that sees the full signal in order to extract each embedding. Hence, in theory, a model based on these features could learn to predict the target class based on the timing of the non-speech regions, which may cause it to diagnose a problem where there is none. 

For our purposes, we need features that can represent background sounds, and channel and reverberation characteristics, without duration information. This can be done with hand-crafted features based on the spectrogram. Such features would be unable to encode duration information since they are extracted from fixed short-term windows over the audio.
Our toolkit implements two such features, though other alternatives could be easily added: raw spectrograms and Mel-Frequency Cepstral Coefficients (MFCC). As an additional precaution, we only extract features over windows that are fully contained within the segment rather than padding the segment at the end of the segment, as is done by default by some libraries. This way, we avoid including features that may flag the end of the segment. For spectrograms, we extract a magnitude STFT using a 400-point FFT and a hop length of 10 ms with 15 ms overlap. For MFCCs, we extract 40 coefficients using the same hop length and overlap. Both feature sets are extracted after resampling the input audio to 16 kHz using the \textit{torchaudio} library.

\subsection{Chunking procedure}\label{sec:chunks}

To prevent the model from being able to encode information about the timing and total duration of non-speech regions, we process the features extracted over each non-speech segment as shown in Figure \ref{fig:pipeline}: the features over the individual segments are concatenated and the resulting sequence is chunked into windows of 5 seconds with a 4-second overlap. Each chunk is treated independently during training and evaluation. This process prevents the classifier from having access to duration information since it only ever sees fixed-length chunks. 

In the experiments below, we also show results for features extracted over speech regions for comparison and benchmarking. In this case, concatenation and chunking is not done. The speech segments are treated as individual instances (commonly called inter-pausal units, IPUs) during modeling and inference.

The feature sequences corresponding to each chunk (for non-speech) or IPU (for speech) are used as independent samples for model training. The target label for each chunk or IPU is inherited from the waveform from which it was extracted. Then, during inference, the scores generated by the model for each chunk or IPU in a waveform are averaged to produce the final score for that waveform. 

\subsection{Classification Model}\label{sec:classifier}
The classifier consists of a 1D Convolutional Neural Network (CNN) with a stride of 2, followed by batch normalization and a ReLU activation. The output is mean pooled over the temporal dimension to obtain a fixed-length representation for each chunk (or IPU). This pooled representation is then passed through a linear projection layer, followed by ReLU and dropout. Finally, a linear layer maps the projected features to the binary output. The number of channels of the CNN and the projection layer can be set through a configuration file.

For experiments using W2V2, the activations from all layers are first averaged over time, and combined into a single vector by a trainable linear layer which is processed by a hidden layer with 64 outputs and the output layer.

Assuming that the number of samples may be relatively small, the training and evaluation process is done with an 8-fold cross-validation approach. The number of training epochs is determined by early stopping with a configurable patience value, using one of the training folds for validation. The final model is trained on all training folds using the number of epochs selected on the validation fold. To assess variability due to the specific choice of folds, we run each system using 10 random seeds. 

\subsection{Metrics}\label{sec:metrics}

As performance metric, we report the area under the ROC curve (AUC), because it allows us to measures how useful the scores are for discriminating one class from the other over all possible thresholds. The toolkit performs a bootstrapping analysis over the test set to provide confidence intervals.

\section{Experiments and Discussion}\label{sec:results}

\begin{figure*}[t]
    \centering
    \includegraphics[width=17cm]{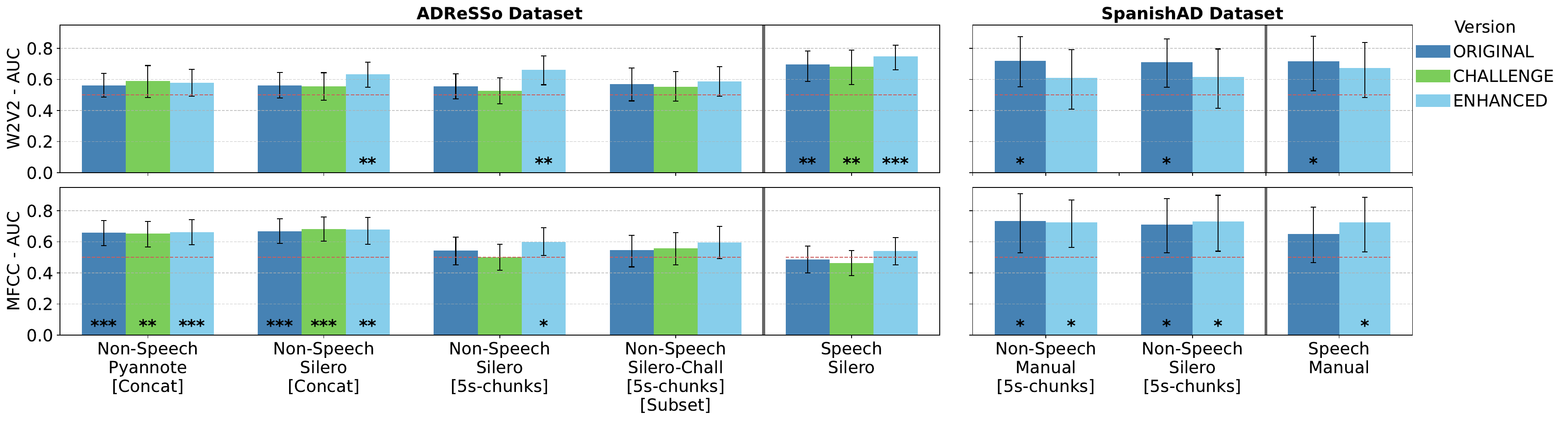}
    \vspace{-0.7cm}
    \caption{Results on the ADReSS$_o$ and SpanishAD datasets across different features (W2V2 and MFCC), pre-processing methods (original, challenge and enhanced), and selected regions (non-speech and speech). Asterisks indicate results that are significantly different from chance (*: $p<0.05$, **: $p<0.01$, ***: $p<0.001$). Error bars correspond to the 5\% to 95\% confidence interval. See Section~\ref{sec:results} for a description of the system names.}
    \vspace{-0.35cm}
    \label{fig:results}
\end{figure*}

To evaluate our proposed approach, we apply our method on two Alzheimer's disease (AD) speech datasets, both based on the Cookie Theft picture description task from the Boston Diagnostic Aphasia Examination. The first is the \textbf{ADReSS$_o$} challenge dataset \cite{luz21_interspeech}, from which we use the training partition only since test labels were not released. The dataset contains 79 control and 87 AD participants, each contributing one recording (mean duration $78$ sec., std $=39$ sec.). We evaluate three versions of the waveforms: enhanced waveforms released for the challenge (CHALLENGE), the original raw recordings without enhancement (ORIGINAL), and a version enhanced using our toolkit (ENHANCED), as described in Section~\ref{sec:speech_enhancement}.

The second corpus is \textbf{SpanishAD} \cite{sanz2022automated}, a Chilean Spanish clinical dataset collected at the Memory and Neuropsychiatry Clinic (Universidad de Chile and Hospital del Salvador), comprising 39 speakers. To balance gender and condition, we retained 32 speakers (16 per class). Recordings have a mean duration of $88.1$ seconds (std $=28.64$). SpanishAD provides high-quality manual inter-pause unit (IPU) alignments, where speech segments are separated by pauses no longer than $0.2$ seconds and non-verbal vocalizations (e.g., coughs, laughs, filled pauses) are labeled as speech content. 

Unfortunately, the acquisition protocol of this dataset introduced spurious correlations. The recording format and sampling rate are strongly associated with the class (76\% AD samples were recorded at $11$ kHz, while only 11\% control samples were recorded at that rate, with all other samples recorded at 44 to 48 kHz). Recordings were obtained in different rooms with different microphones and no control over the acoustic environment. 
Notably, metadata alone (e.g., codec/format, sampling rate, and duration), extracted with \texttt{ffprobe} (FFmpeg), was sufficient to predict patient condition using a Random Forest model (AUC $=0.72$). 
In our experiments, we down-sampled all waveforms to $11.025$ kHz before any further processing, and compared performance on these recordings (ORIGINAL), and after denoising with our toolkit (ENHANCED). 
As we will see, even after resampling and enhancement, the target class can still be predicted from the acoustic conditions. 

Using the manual annotations available in the SpanishAD dataset as a reference, we benchmarked several VAD systems. Silero VAD \cite{SileroVAD} with a speech probability threshold of $0.2$ resulted in a good trade-off between error rates for our purposes, with a speech leakage of $2.37\%$ and missed non-speech of $22\%$. This setup prioritizes a low speech leakage at the expense of a relatively high rate of missed non-speech, which is much less problematic for our purpose. 
Other VAD systems resulted in higher rates of missed non-speech for a similar rate of speech leakage, indicating that, for this data, Silero VAD was the best option. This conclusion, though, may not hold for other datasets, so we recommend doing this analysis using a subset of annotated samples from the dataset of interest.


The Figure~\ref{fig:results} shows the results for the ADReSS$_o$ and SpanishAD datasets using MFCC and W2V2 features. We include W2V2 results for comparison, though we do not recommend their use for an actual diagnostic analysis (see Section \ref{sec:features}). We exclude spectrogram results since they were worse than those for MFCCs for both datasets. We note, though, that the best feature type for this analysis may depend on the type of artifact present in the data. Hence, we recommend trying both MFCC and spectrogram for any new dataset.

We compare systems trained and evaluated only on speech or non-speech regions. Speech and non-speech regions are obtained using different VAD systems and, for SpanishAD, manual annotations. In ADReSS$_o$, speech regions include segments from both the participant and the experimenter, while in SpanishAD they include only the participant's speech. Bars indicate the mean performance across 10 random seeds, and uncertainty is estimated via 1000 bootstrap resamples per seed. 

The systems called Non-Speech correspond to  four different approaches for extracting and modeling the non-speech regions. The ones with [Concat] in their name model the features (over the non-speech regions obtained with Silero or Pyannote) after concatenation but without the chunking process explained in Section \ref{sec:chunks}. Hence, the model has access to the full sequence during training and inference. The results for these systems are significantly above chance for the MFCC features for both VAD systems suggesting that a spurious correlation is present. 

Yet, these systems have access to the duration of the waveform and may be able to leverage this information for making the prediction, since the duration distribution is markedly different between AD and control samples. 
Notably, this same problem is present in the system proposed in prior work \cite{liu2024cleverhans}, where they report an above-chance accuracy of 0.61 for a system based on W2V2 fine-tuned using the concatenated audio from all non-speech regions. 

To control for this potential leakage of legitimate task-dependent information, we propose the chunking approach described in Section \ref{sec:chunks}, named [5s-chunks] in the figure. In this case the model never sees the full feature sequence, preventing it from learning the duration of the samples. 
When using this approach, the performance on the original and challenge samples drops to random for both feature sets, but stays significantly different from chance for the enhanced samples. We hypothesized that this was due to leakage of speech into the non-speech regions for this data, perhaps due to the distortions introduced by the enhancement model which cause a mismatch for the VAD model (though enhanced signals sound clean to the ear, as can be appreciated in the samples provided in the repository).

To address this issue, we run the system called Silero-Chall [5s chunks] [Subset] where the VAD is run on the challenge samples and applied to the other two cases. Further, samples are audited to discard cases where there is speech leakage (29 out of 166 samples were discarded). This is the only system in which speech leakage is fully controlled. In this case, results are not significantly different from chance for any of the datasets, though the result for the enhanced signals strongly suggest that a small signal to detect the target is, in fact, present in the non-speech regions, and is emphasized by the enhancement process.

Comparing the results over the speech and non-speech regions, we can see that, while MFCC tends to give better results than W2V2 for non-speech regions, the trend is reversed over speech regions. This is a reasonable result given that the W2V2 model was trained on speech data with little amount of silence. Notably, our hypothesis that W2V2 might be able to model segment duration information is not supported by the results, since these features do not lead to better performance than MFCCs on non-speech regions. Yet, note that this conclusion may differ on other datasets. For this reason, we recommend never using contextualized features like W2V2 for this analysis.

The plot on the right of Figure \ref{fig:results}, shows results on the SpanishAD dataset. In this case, results are significantly above chance for both the original and the enhanced signals. 
When comparing automatically derived non-speech regions with those obtained from manual annotations, results are similar though with a slight increase in AUC for the Pyannote-based VAD, suggesting that some speech may be leaking into the non-speech regions.
Notably, speech enhancement does not mitigate the problem. Whatever artifacts were caused by the collection do not appear to be reduced by the enhancement process, despite the fact that, perceptually, the audio is greatly improved.
Interestingly, the speech-based results are not better than those obtained over non-speech regions, highlighting the severity of the spurious correlations present in this dataset.

\section{Conclusions}

We propose a method for uncovering spurious correlations from speech datasets labeled with some speech-related class, like a patient condition, emotion or speaker identity. The method attempts to detect the target class based on the non-speech regions of the signal. Results significantly better than random indicate that the recording conditions correlate with the target class. Such datasets should not be used for speech-based analysis, except by using very high-level or manually extracted features that would not be able to encode the recording conditions, like those based on transcriptions or carefully-curated pitch and timing features. It is important to note that a result not significantly better than random does not imply that spurious correlations do not exist. Many other factors may cause random results even when spurious correlations do exist, like a number of samples that is insufficient for the model to learn, samples with too little non-speech content, or the fact that some artifacts may only occur over the speech regions.

We release a toolkit implementing the proposed method, which can be used on any speech dataset with waveform-level speech-related labels and sufficient non-speech content.
The analysis can also be applied during data collection, enabling adjustments to the protocol if a design issue is detected through this method.

\section{Acknowledgements}
We gratefully acknowledge the support of NVIDIA Corporation for the donation of a Titan Xp GPU. Adolfo García is supported by GBHI, Alzheimer’s Association, and Alzheimer’s Society (Alzheimer’s Association GBHI ALZ UK-22-865742), as well as ANID (FONDECYT Regular 1210176). This work was partially supported by the Air Force Office of Scientific Research, United States, under award numbers FA9550-18-1-0026 and F9550-21-1-0445, and by the European Union’s Horizon 2020 research and innovation programme under the Marie Skłodowska-Curie grant agreement No 101007666.

\section{Generative AI Use Disclosure}
We used a generative AI tool for light language editing and translation. All experimental design, implementation decisions, analyses, and interpretations were carried out and validated by the authors, who take full responsibility for the work.

\bibliographystyle{IEEEtran}
\bibliography{mybib}

\end{document}